\documentclass[preprint]{aastex}
\usepackage{natbib, aas_macros}
\shorttitle{CO selective dissociation in $\rho$ Oph A}
\shortauthors{Yamagishi, M. et al.}

\begin{document}

\title{ALMA observations of layered structures due to CO selective dissociation in the $\rho$ Ophiuchi A plane-parallel PDR}

\author{M. Yamagishi\altaffilmark{1}, C. Hara\altaffilmark{2}, R. Kawabe\altaffilmark{2,3,4}, F. Nakamura\altaffilmark{2,3,4}, T. Kamazaki\altaffilmark{3}, T. Takekoshi\altaffilmark{5}, Y. Shimajiri\altaffilmark{6}, H. Nomura\altaffilmark{7}, S. Takakuwa\altaffilmark{8}, J. Di Francesco\altaffilmark{9}}
\email{yamagish@ir.isas.jaxa.jp}

\altaffiltext{1}{Institute of Space and Astronautical Science, Japan Aerospace Exploration Agency, Chuo-ku, Sagamihara, Kanagawa 252-5210, Japan}
\altaffiltext{2}{Graduate School of Science, The University of Tokyo, Bunkyo-ku, 113-0033, Tokyo, Japan}
\altaffiltext{3}{National Astronomical Observatory of Japan, National Institutes of Natural Sciences, 2-21-1 Osawa, Mitaka, Tokyo 181-8588, Japan}
\altaffiltext{4}{The Graduate University for Advanced Studies (SOKENDAI), 2-21-1 Osawa, Mitaka, Tokyo 181-0015, Japan}
\altaffiltext{5}{Institute of Astronomy, The University of Tokyo, 2-21-1 Osawa, Mitaka, Tokyo 181-0015, Japan}
\altaffiltext{6}{Laboratoire AIM, CEA/DSM-CNRS-Universit\'e Paris Diderot, IRFU/Service d'Astrophysique, CEA Saclay, F-91191 Gif-sur-Yvette, France}
\altaffiltext{7}{Department of Earth and Planetary Sciences, Tokyo Institute of Technology, 2-12-1 Ookayama, Meguro, Tokyo, 152-8551, Japan}
\altaffiltext{8}{Department of Physics and Astronomy, Graduate School of Science and Engineering, Kagoshima University, 1-21-35 Korimoto, Kagoshima, Kagoshima 890-0065, Japan}
\altaffiltext{9}{NRC Herzberg Astronomy and Astrophysics Research Centre, 5071 W Saanich Rd, Victoria BC V9E 2E7, British Columbia, Canada}

\begin{abstract}
We analyze $^{12}$CO($J$=2--1), $^{13}$CO($J$=2--1), C$^{18}$O ($J$=2--1), and 1.3~mm continuum maps of the $\rho$ Ophiuchi A photo-dissociation region (PDR) obtained with ALMA.
Layered structures of the three CO isotopologues with an angular separation of 10$\arcsec$ = 6.6$\times$10$^{-3}$~pc = 1400~au are clearly detected around the Be star, S1 (i.e., each front of emission shifts from the near to far side in order of $^{12}$CO, $^{13}$CO, and C$^{18}$O).
We estimate the spatial variations of $X$($^{13}$CO)/$X$(C$^{18}$O) abundance ratios, and find that the abundance ratio is as high as 40 near the emission front, and decreases to the typical value in the solar system of 5.5 in a small angular scale of 4$\arcsec$ = 2.6$\times$10$^{-3}$~pc = 560~au.
We also find that the $I$($^{12}$CO(2--1))/$I$($^{13}$CO(2--1)) intensity ratio is very high ($>$21) in the flat-spectrum young stellar object, GY-51, located in the PDR.
The enhancement of the ratios indicates that the UV radiation significantly affects the CO isotopologues via selective dissociation in the overall $\rho$ Ophiuchi A PDR, and that the $\rho$ Ophiuchi A PDR has a plane-parallel structure.
\end{abstract}

\keywords{submillimeter: ISM --- ISM: photo-dissociation region ---  ISM: molecules --- individual object ($\rho$ Ophiuchi A)}

\section{Introduction}

The photo-dissociation region (PDR) is defined by a region where heating and chemistry are regulated by far-ultraviolet (FUV: 6--13.6~eV) photons (\citealt{Tielens85, Hollenbach91}).
One of the big outstanding issues in current PDR study is that there is no unified understanding about the PDR spatial structure.
The ideal plane-parallel PDR model proposed by \citet{Tielens85} has a homogeneous density structure, and shows a plane-parallel PDR structure in between HII region and molecular cloud.
Recently reported observational results, however, have not been explainable with a simple plane-parallel PDR model.
For example, [CI] and $^{13}$CO emission has been observed to be coincident in high-mass star-forming regions (Orion Bar, M17, S160, and W51; \citealt{Keene85, Genzel88, Tauber95, Shimajiri13}).
Since [CI] and $^{13}$CO are probes of PDR and molecular gas, respectively, non-coincident spatial distributions are expected in the plane-parallel PDR.
To interpret such inconsistency, the alternative ideal model was proposed (clumpy PDR model; \citealt{Stutzki88, Meixner93, Spaans96}), which has inhomogeneous density structures and shows mixed spatial distributions of PDRs and molecular gas.
To distinguish between the two PDR models, a key parameter is spatial resolution.
If the spatial resolution is low, the spatial distributions of the PDR and molecular cloud can appear coincident even in the plane-parallel PDR.
Hence, observations with the high spatial resolution are necessary to investigate PDR models.

One of the expected phenomena in PDRs is CO selective dissociation (\citealt{Glassgold85, Yurimoto04}).
In this process, UV radiation selectively dissociates less-abundant CO isotopologues more effectively than more-abundant ones because of the difference in self-shielding.
As a result of CO selective dissociation, CO isotopologues are expected to exhibit layered structures at the PDR interface from more- to less-abundant isotopologues as a function of distance from the excitation star.
It is difficult to form such layered structures in a clumpy PDR because, due to the inhomogeneous density, the penetration depth of UV radiation does not simply depend on the distance from the exciting star.
Hence, if layered structures of CO isotopologues are detected around an excitation star, the spatial structure of the PDR may be plane-parallel.
In recent years, CO selective dissociation has been studied for nearby molecular clouds with the Nobeyama 45m telescope  (Orion: \citealt{Shimajiri14}; L1551: \citealt{Lin16}).
They observed the molecular clouds overall, and found large-scale ($\sim$0.2~pc) spatial variations of $X$($^{13}$CO)/$X$(C$^{18}$O) abundance ratios between the surfaces and insides of the molecular clouds.
The $X$($^{13}$CO)/$X$(C$^{18}$O) abundance ratios are in ranges of 8--30 and 4--20 in Orion and L1551, respectively. 
CO layered structures around the excitation star, however, have not been directly detected yet.
Hence, observations with higher spatial resolution ($<$0.05~pc) are required to examine PDR structures via CO selective dissociation.

\begin{figure}
\epsscale{0.65}
\plotone{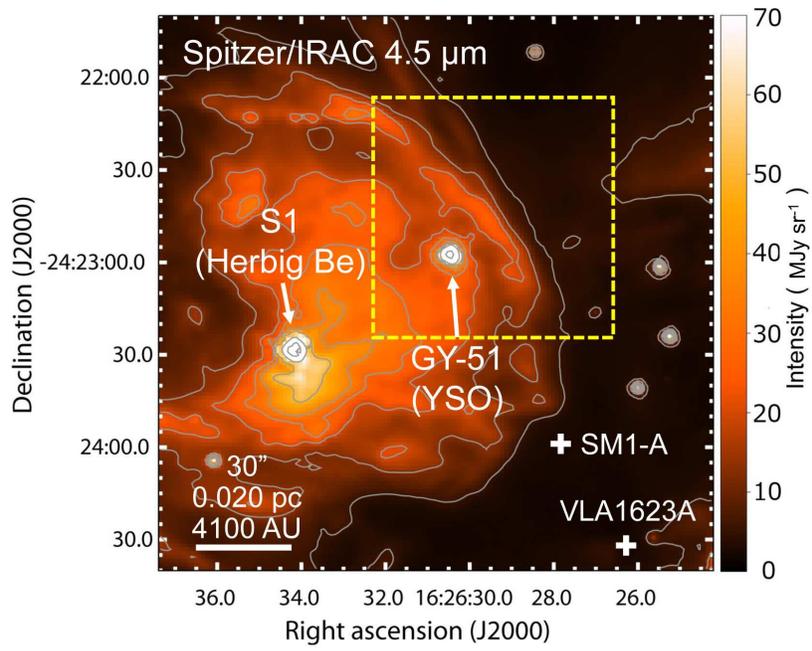}
\caption{$Spitzer$/IRAC 4.5~$\micron$ map of $\rho$ Oph A. The yellow dashed box indicates the target area in the present study. The contour levels are 5, 12.5, 20, 27.5, 35, 50, 100, and 400 MJy~sr$^{-1}$. Crosses show positions of SM1-A (\citealt{Kamazaki01}) and VLA1623A as positional references.}
\label{obsarea}
\end{figure}

The main motivation of the present study is examining the PDR structure using the high spatial resolution of ALMA.
To achieve this goal, we focus on the $\rho$ Ophiuchi A ($\rho$ Oph A) region, which is the closest PDR ($d$=137.3$\pm$1.2~pc; \citealt{Ortiz-Leon17}), and is illuminated by the Herbig Be star, S1.
Figure~\ref{obsarea} shows a $Spitzer$/IRAC 4.5~$\micron$ image of the $\rho$ Oph A PDR, which shows a clear shell-like structure.
Such a structure was also detected in H$_2$S(4) and [OI] emission (\citealt{Larsson17}).
The FUV intensity at the PDR shell was estimated to be $G_0$=(3--4)$\times$10$^3$ (\citealt{Shimajiri17}).
In addition, bright C$^{18}$O, N$_2$H$^+$, and dust continuum emission was detected behind the PDR shell with the James Clerk Maxwell Telescope (JCMT) and the Atacama Pathfinder Experiment (\citealt{Motte98, Liseau15, Larsson17}).
Hence, $\rho$ Oph A is the closest typical PDR adjacent to an HII region and molecular cloud, which is ideal for the present study.

\section{Observation and data reduction}

We analyzed ALMA Cycle-2 data at Band-6 for a 2$\arcmin$$\times$3$\arcmin$ region of $\rho$ Oph A obtained using the 12-m Array and the ACA 7-m Array (PI: F. Nakamura; Project code: 2013.1.00839.S), which covered 1.3 mm (226~GHz) continuum and the $^{12}$CO, $^{13}$CO, C$^{18}$O ($J$=2--1) lines.
The uv ranges sampled in the 12-m Array and 7-m Array data were 12.5--348 k$\lambda$ and 8.1--48 k$\lambda$, respectively ($\lambda\sim$1.3~mm).
The minimum uv distance of the combined data corresponds to 25$\arcsec$.
The images used in the present paper were reduced using the same methods described in \citet{Kawabe18}.
We focus on a 1$\farcm$3$\times$1$\farcm$3 region associated with the shell structure of $\rho$ Oph A in Fig.~\ref{obsarea}.
We used a common synthesized beam (1$\farcs$45 $\times$ 0$\farcs$94) and velocity grid  (0.2~km~s$^{-1}$) in the map making.
The beam size corresponds to 9.6$\times$10$^{-4}$~pc $\times$ 6.3$\times$10$^{-4}$~pc = 200~au $\times$ 130~au at the distance of $\rho$ Oph A, which is 40 and 20  times higher spatial resolution than that of previous single-dish observations for Orion and L1551, respectively (\citealt{Shimajiri14, Lin16}).
The 1$\sigma$ noise levels of the $^{12}$CO($J$=2--1), $^{13}$CO ($J$=2--1), and C$^{18}$O ($J$=2--1) maps are 45~mJy~beam$^{-1}$, 45~mJy~beam$^{-1}$, and 33~mJy~beam$^{-1}$, respectively, at a velocity resolution of 0.2~km~s$^{-1}$, while that of the continuum map is 0.24~mJy~beam$^{-1}$.

\section{Results \& discussion}
\subsection{$\rho$ Oph A PDR}

\begin{figure}
\epsscale{1.0}
\plotone{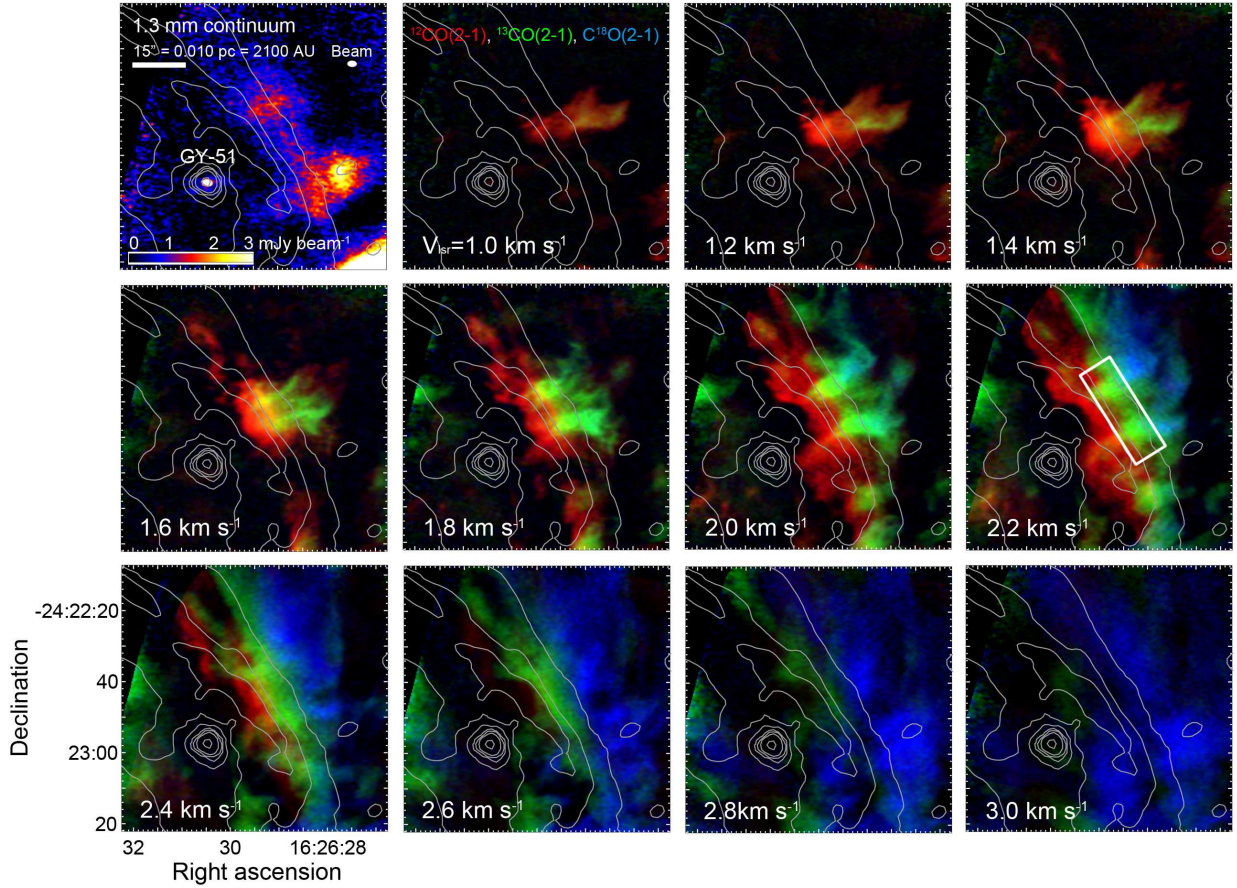}
\caption{1.3~mm continuum map and color-composite channel maps of $^{12}$CO(2--1) (red), $^{13}$CO(2--1) (green), and C$^{18}$O(2--1) (blue) emissions at $V_{\mathrm{lsr}}$=1.0--3.0 km~s$^{-1}$ for the yellow dashed box in Figure~\ref{obsarea}. Contours are the same as in Fig.~\ref{obsarea}. The color range in the 1.3~mm continuum map is 0--3~mJy~beam$^{-1}$, while the peak intensity is 37.7~mJy~beam$^{-1}$ associated with GY-51. The color range is common among the channel maps and is linear from 0~\% to 80\% of the peak intensity in this area ($^{12}$CO: 4.98~Jy~beam$^{-1}$; $^{13}$CO: 2.43~Jy~beam$^{-1}$; C$^{18}$O: 1.32~Jy~beam$^{-1}$).}
\label{channel_map}
\end{figure}

\begin{figure}
\epsscale{0.5}
\plotone{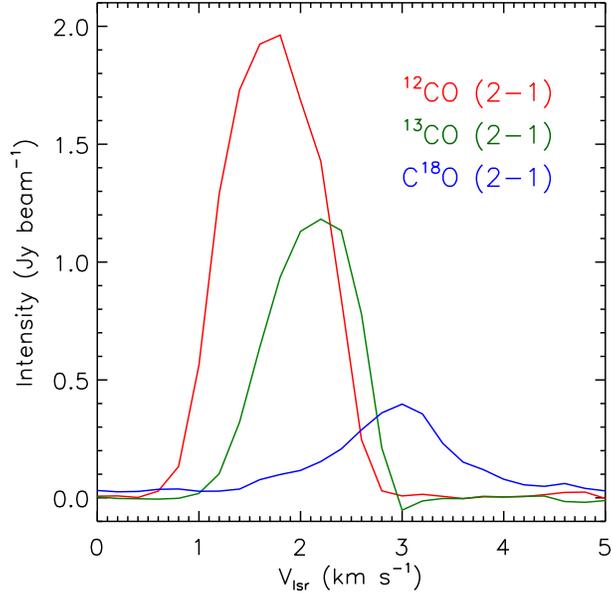}
\caption{Averaged spectra of $^{12}$CO(2--1) (red), $^{13}$CO(2--1) (green), and C$^{18}$O(2--1) (blue) emissions at the layered structure (the white-box region in Fig.~\ref{channel_map}, $V_{\mathrm{lsr}}$=2.2 km~s$^{-1}$).}
\label{shell_spec}
\end{figure}

Figure~\ref{channel_map} shows channel maps of $^{12}$CO ($J$=2--1), $^{13}$CO ($J$=2--1), and C$^{18}$O ($J$=2--1) emission between 1.0~km~s$^{-1}$ and 3.0~km~s$^{-1}$ and the 1.3~mm continuum map tracing optically-thin dust continuum emission.
Extended CO structures are detected along the PDR shell at $V_{\mathrm{lsr}}>$1.8 km~s$^{-1}$, while compact structures are detected on the PDR shell at $V_{\mathrm{lsr}}$=1.0--1.6 km~s$^{-1}$.
The latter have been reported as point-like sources in $^{12}$CO ($J$=3--2) (\citealt{Kamazaki03}), which may be a molecular-gas outflow through a hole in the PDR shell.
It is notable that distinct layered structures of the CO isotopologues are clearly detected at 2.2 km~s$^{-1}$, 2.4 km~s$^{-1}$ and 2.6 km~s$^{-1}$; $^{12}$CO, $^{13}$CO, and C$^{18}$O emission is bright inside the shell, on the shell, and behind the shell, respectively.
The layered structures indicate that the UV radiation from S1 efficiently dissociates CO isotopologues in order of C$^{18}$O, $^{13}$CO, and $^{12}$CO due to selective dissociation.
Such layered structures are observed when the CO isotopologues are homogeneously distributed in the PDR.
Therefore, $\rho$ Oph A is interpreted as a plane-parallel-like PDR (i.e. homogeneous-density PDR) rather than a clumpy-like PDR.
The separation between the layers is $\sim$10$\arcsec$ (= 6.6$\times$10$^{-3}$~pc = 1400~au), which is not spatially resolved using current large single-dish telescopes.
Therefore, detailed study of CO selective dissociation is uniquely visible with ALMA.
At the surface of the Orion Bar, \citet{Goicoechea17} reported layered structures of reactive ion emission with ALMA, which is a complementary result in terms of probes used to examine such structures.
The systemic velocity of $\rho$ Oph A was measured to be 3.7 km~s$^{-1}$ in previous N$_2$H$^+$ observations (\citealt{Mardones97}).
We, however, find that $^{12}$CO and $^{13}$CO emission is absent near the systemic velocity.
For example, Fig.~\ref{shell_spec} presents spectra of the CO isotopologues at the layered structure, which shows that the $^{12}$CO and $^{13}$CO emission is attenuated at $V_{\mathrm{lsr}}>$2.0~km~s$^{-1}$ and $>$2.4~km~s$^{-1}$, respectively.
Presumably, the $^{12}$CO and $^{13}$CO emission has been significantly attenuated due to line-of-sight absorption and/or resolved out around the systemic velocity by spatial filtering.

We verify that UV emission significantly affects CO molecules.
The peak $^{12}$CO intensity associated with the PDR shell ($V_{\mathrm{lsr}}$ $>$ 1.8 km~s$^{-1}$) is 3.8 Jy~beam$^{-1}$ which corresponds to a brightness temperature of 69~K.
The gas temperature heated by UV radiation is generally described as $G_0 = (T_{\mathrm{gas}}/12.2)^5$ (\citealt{Hollenbach91}), assuming thermal equilibrium between gas and dust.
By using the UV intensity of $G_0$=(3--4)$\times$10$^3$ (\citealt{Shimajiri17}), the expected gas temperature is $T_{\mathrm{gas}} = 64~(G_0/4\times10^3)^{1/5}$~K, which is roughly consistent with the observed brightness temperature.
This consistency indicates the heating of CO isotopologues by the UV radiation around the PDR shell.

Next, we examine spatial variations of $X$($^{13}$CO)/$X$(C$^{18}$O) abundance ratios ($R_{13/18}$) for the CO layered structures at $V_{\mathrm{lsr}}$= 2.2 km~s$^{-1}$, 2.4 km~s$^{-1}$, and 2.6 km~s$^{-1}$.
By the definition of $X$($^{13}$CO) = $N$($^{13}$CO)/$N$(H$_2$) and $X$(C$^{18}$O) = $N$(C$^{18}$O)/$N$(H$_2$), $R_{13/18}$ is equivalent to the $N$($^{13}$CO)/$N$(C$^{18}$O) column density ratio.
We estimated the excitation temperature ($T_{\mathrm{ex}}$) of CO using the peak brightness temperature of $^{12}$CO in $V_{\mathrm{lsr}}>$1.8 km~s$^{-1}$ and the optical depth of $^{13}$CO ($\tau_{13}$) and C$^{18}$O ($\tau_{18}$) using that $T_{\mathrm{ex}}$.
Here we assumed that the excitation temperature is common among the CO isotopologues, and that the $^{12}$CO ($J$=2--1) emission is heavily optically thick.
Finally, $N$($^{13}$CO) and $N$(C$^{18}$O) were estimated under the LTE condition for simplicity by using equation (11) in \citet{Nishimura15}, as only single transitions of the CO isotopologues are available.
$R_{13/18}$ estimated under LTE and non-LTE assumptions are coincident within a factor of $\sim$2 in the massive star-forming region G29.96-0.02 (\citealt{Paron18}).
In our estimation, $N$($^{13}$CO) and $N$(C$^{18}$O) are not reliably determined behind the PDR shell and near the systemic velocity ($V_{\mathrm{lsr}}$=3.7 km~s$^{-1}$).
In such areas, the $^{12}$CO emission is significantly attenuated due to absorption and/or spatial filtering (see also Fig.~\ref{shell_spec}), which cause underestimates of $T_{\mathrm{ex}}$.
Hence, the resulting estimations of $N$($^{13}$CO) and $N$(C$^{18}$O) are unreliable due to the unusually small $T_{\mathrm{ex}}$ relative to that expected from the observed $^{13}$CO and C$^{18}$O emission.
Hence, we set the lower limit of the excitation temperature as 45~K, as previously estimated from the JCMT $^{12}$CO ($J$=3--2) data (\citealt{White15}), which is not affected by spatial filtering.
The resulting ranges of $T_{\mathrm{ex}}$, $\tau_{13}$, $\tau_{18}$,  $N$($^{13}$CO), and $N$(C$^{18}$O) are 45--75~K, 0.03--4.4, 0.02--0.77, 1.6$\times$10$^{16}$--1.4$\times$10$^{18}$ cm$^{-2}$, and 1.1$\times$10$^{16}$--2.4$\times$10$^{17}$ cm$^{-2}$, respectively.

\begin{figure}
\epsscale{1.0}
\plotone{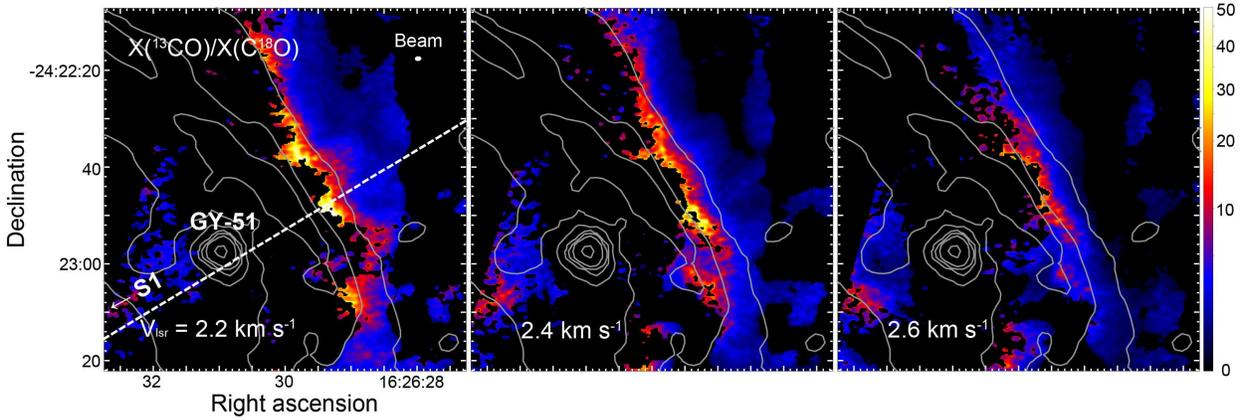}
\caption{$X$($^{13}$CO)/$X$(C$^{18}$O) abundance ratio maps at $V_{\mathrm{lsr}}$=2.2--2.6~km~s$^{-1}$. Contours are the same as in Fig.~\ref{obsarea}. The dashed line indicates the direction used to calculate the abundance ratio profiles in Fig.~\ref{profile}.}
\label{ratio}
\end{figure}

Figure~\ref{ratio} shows the $R_{13/18}$ maps at $V_{\mathrm{lsr}}$= 2.2 km~s$^{-1}$, 2.4 km~s$^{-1}$, and 2.6 km~s$^{-1}$.
Each map shows a single gradient structure in the radial direction from S1.
$R_{13/18}$ is as high as 40 near the shell, which is significantly larger than the 5.5 seen in the solar system (\citealt{Anders89}).
Hence, $R_{13/18}$ appears to have been enhanced by the UV radiation.
We also checked $R_{13/18}$ under assumptions of $T_{\mathrm{ex}}>$50~K and 55~K, and confirmed that the overall trend in Figure~\ref{ratio} does not change, although higher $T_{\mathrm{ex}}$ generally shows lower $R_{13/18}$ by $\sim$20~\%.
Therefore, our results do not depend on the assumptions of $T_{\mathrm{ex}}$ and thus $\tau_{13}$ or $\tau_{18}$. 
The maps at lower $V_{\mathrm{lsr}}$ show higher $R_{13/18}$.
Indeed, variations among different $V_{\mathrm{lsr}}$ are consistent with expected degrees of absorption and spatial filtering; emission near the systemic velocity is heavily attenuated, and $^{13}$CO is more sensitive to such effects than C$^{18}$O.
Hence, the observed $R_{13/18}$ should be treated as the lower limit to the actual $R_{13/18}$.

\begin{figure}
\epsscale{0.65}
\plotone{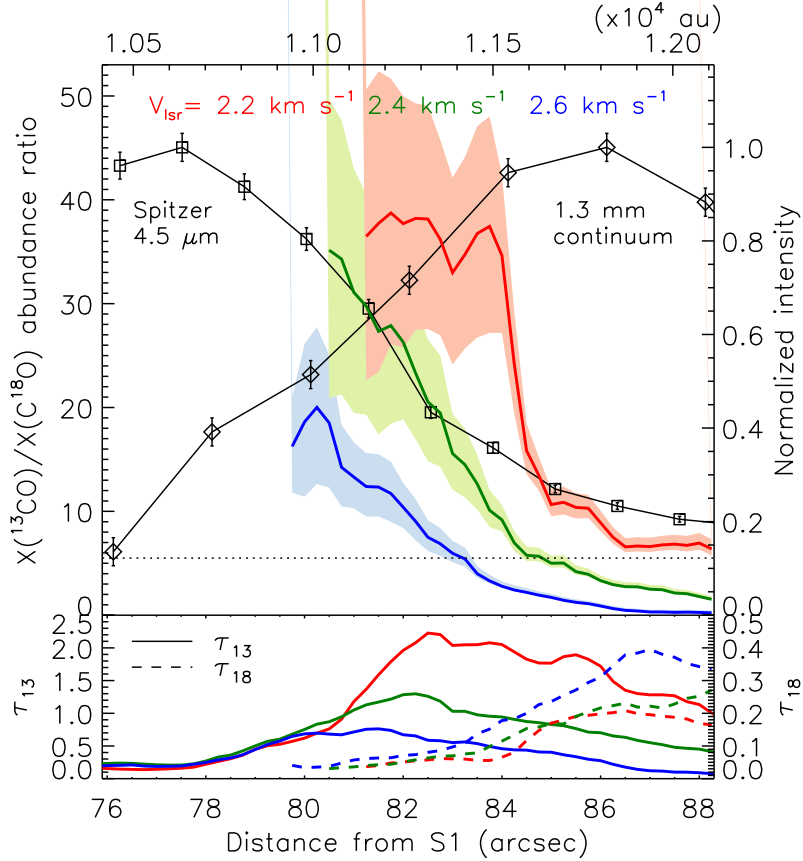}
\caption{(top) Spatial variations of the $X$($^{13}$CO)/$X$(C$^{18}$O) abundance ratio at $V_{\mathrm{lsr}}$=2.2--2.6~km~s$^{-1}$ (color-corded lines; left axis) and those of the $Spitzer$ 4.5~$\micron$ (squares; right axis) and 1.3~mm continuum intensities (diamonds; right axis) as a function of the distance from S1. These are calculated from averaged profiles in a rectangular region with a width of 1$\farcs$5 along the dashed line in Fig.~\ref{ratio}. The $Spitzer$ 4.5~$\micron$ and 1.3~mm continuum data are normalized by the maximum values of 23.9~MJy~sr$^{-1}$ and 1.0~mJy~beam$^{-1}$, respectively. The dotted line indicates the $X$($^{13}$CO)/$X$(C$^{18}$O) abundance ratio of 5.5 in the solar system (\citealt{Anders89}). Note that the $X$($^{13}$CO)/$X$(C$^{18}$O) abundance ratios smaller than 5.5 in $V_{\mathrm{lsr}}$=2.4 and 2.6 km~s$^{-1}$ may be artificial due to absorption and/or spatial filtering of $^{13}$CO. (bottom) Spatial variations of $\tau_{13}$ and $\tau_{18}$ as functions of the distance from S1. Solid and dashed lines indicate $\tau_{13}$ and $\tau_{18}$, respectively. The different color shows different $V_{\mathrm{lsr}}$ as in the top panel.}
\label{profile}
\end{figure}

Figure~\ref{profile} shows spatial variations of $R_{13/18}$ as a function of projected distance from S1.
The uncertainties of the variations ($e_{13/18}$) are calculated as
\begin{equation}
e_{13/18} = \sqrt{ {\left( \frac{\partial R_{13/18}}{\partial I_{12}} \right)}^2 \sigma_{12}^2 + { \left( \frac{\partial R_{13/18}}{\partial I_{13}} \right)}^2 \sigma_{13}^2 + { \left( \frac{\partial R_{13/18}}{\partial I_{18}} \right)}^2 \sigma_{18}^2},
\end{equation}
where $I_{12}$ ($I_{13}$, $I_{18}$) are the $^{12}$CO ($^{13}$CO, C$^{18}$O) intensity, and $\sigma_{12}$ ($\sigma_{13}$, $\sigma_{18}$) is the 1$\sigma$ uncertainty of $I_{12}$ ($I_{13}$, $I_{18}$).
We find that the $R_{13/18}$ abundance ratio rapidly decreases from 20--40 to the standard value in a small angular scale of 3$\arcsec$--5$\arcsec$, which corresponds to (2.0--3.4)$\times10^{-3}$~pc  = 410--650~au.
Therefore, heavy enhancement of $R_{13/18}$ is limited to the thin layer of the PDR shell. 
As shown in the bottom panel of Fig.~\ref{profile}, $\tau_{13}$ is high ($\sim$1--2) at the surface of the molecular cloud.
Hence, the actual $R_{13/18}$ value at the surface may be higher than that in the present estimate.
The peak positions of $R_{13/18}$ are 80$\arcsec$--82$\arcsec$ = (1.10--1.15)$\times$10$^4$ au away from S1, and are located between the peaks of $Spitzer$ 4.5~$\micron$ and 1.3~mm continuum emissions, each probing a PDR shell and molecular cloud, respectively.
The complementary spatial profiles support the idea that the selective dissociation is dominant at the surface of the molecular cloud.
Note that the $R_{13/18}$ values smaller than 5.5 at $V_{\mathrm{lsr}}$=2.4 and 2.6 km~s$^{-1}$ may be artificial due to absorption and/or spatial filtering of $^{13}$CO emission.

\citet{vanDishoeck88} and \citet{Visser09} calculated the spatial variations of $R_{13/18}$.
\citet{vanDishoeck88} showed that it can be as high as 10, and can decrease to the standard value after an extinction of $\Delta A_V$=0.6~mag ($\Delta d$= 0.4~pc) assuming a hydrogen density of $n_\mathrm{H}=10^3$~cm$^{-3}$, a temperature of 50~K, and a UV intensity of $G_0$=1.7 (UV strength of 1 in units of the Draine radiation field; \citealt{Draine78}).
By simply scaling the model calculation, the expected depth from the peak of $R_{13/18}$ to the standard value is expressed as a function of the density: $\Delta d = 0.4$ ($10^3$~cm$^{-3}/n_\mathrm{H}$)~pc.
The hydrogen density derived for a region 0.1~pc away from the $\rho$ Oph A core is $n_\mathrm{H}\sim10^{5}$~cm$^{-3}$ (\citealt{Loren83}).
Therefore, an equivalent depth of $\Delta d \sim4\times10^{-3}$~pc is expected for the $\rho$ Oph A PDR, which is roughly consistent to the observed value of (2.0--3.4)$\times10^{-3}$~pc.
Hence, our results are consistent to the prediction by \citet{vanDishoeck88}.
\citet{Visser09} presented spatial variations of $R_{13/18}$ under various conditions of a translucent cloud ($n_\mathrm{H}=10^{2-3}$~cm$^{-3}$, $T_{\mathrm{gas}}$=15--100~K, $G_0$=1.7--17).
In their calculation, $R_{13/18}$ was enhanced as high as 60 under gas temperature of 50~K.
Hence, $R_{13/18}$ enhancement up to 20--40 is likely to be possible depending on the parameters.

The layered structures of the CO maps and the gradients seen in $R_{13/18}$ maps consistently indicate that the CO isotopologues are significantly affected by selective dissociation in a plane-parallel PDR.
The present study shows the importance of ALMA for studies of PDRs, as it can spatially resolve the thin layers at the surface of the PDR and directly distinguish a plane-parallel PDR from a clumpy PDR.
By utilizing characteristics of the plane-parallel PDR, the $\rho$ Oph A PDR may be a suitable location for various future studies of PDR chemistry and dynamics.

\subsection{GY-51}

\begin{figure}
\epsscale{0.42}
\plotone{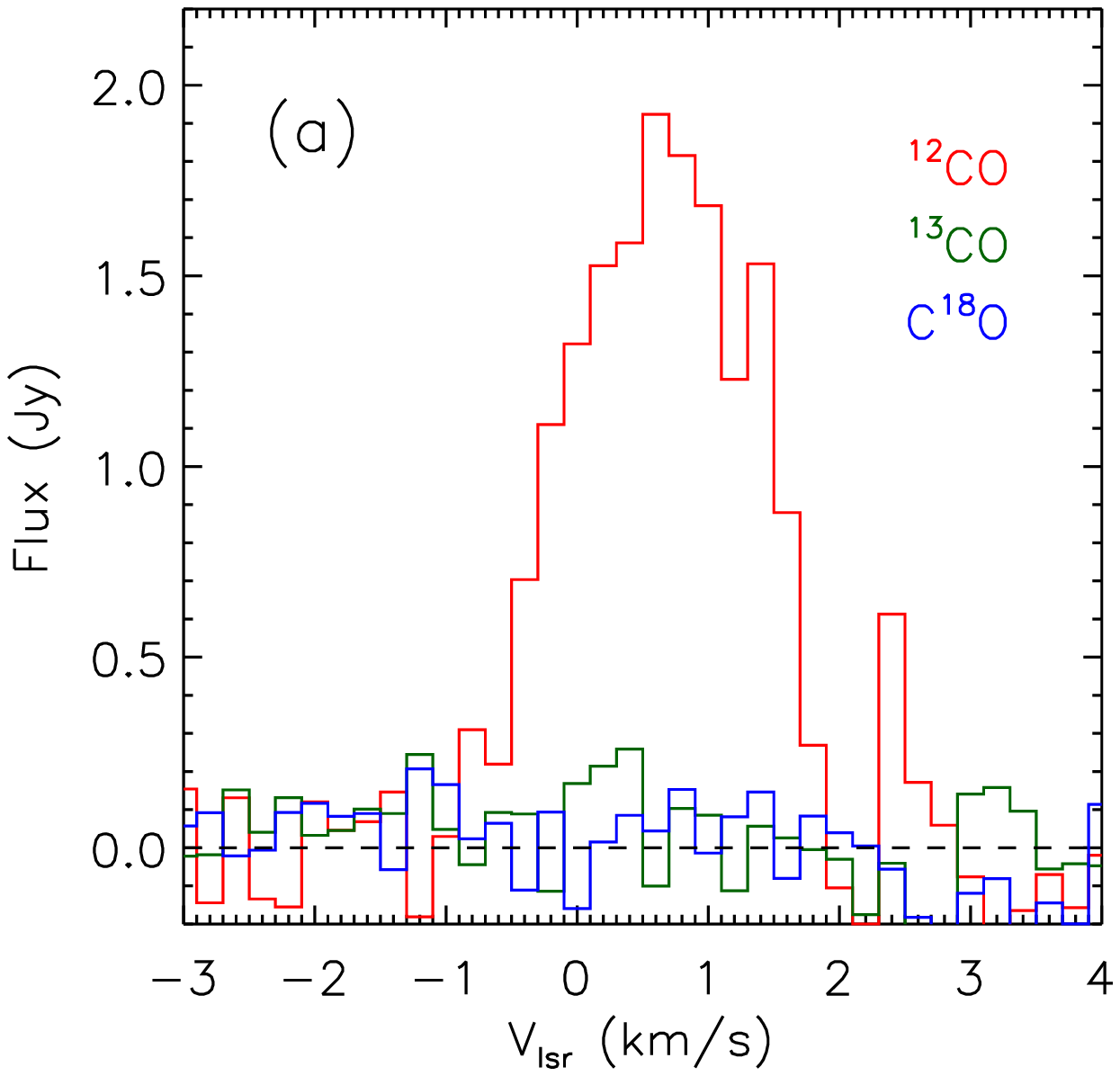}
\epsscale{0.50}
\plotone{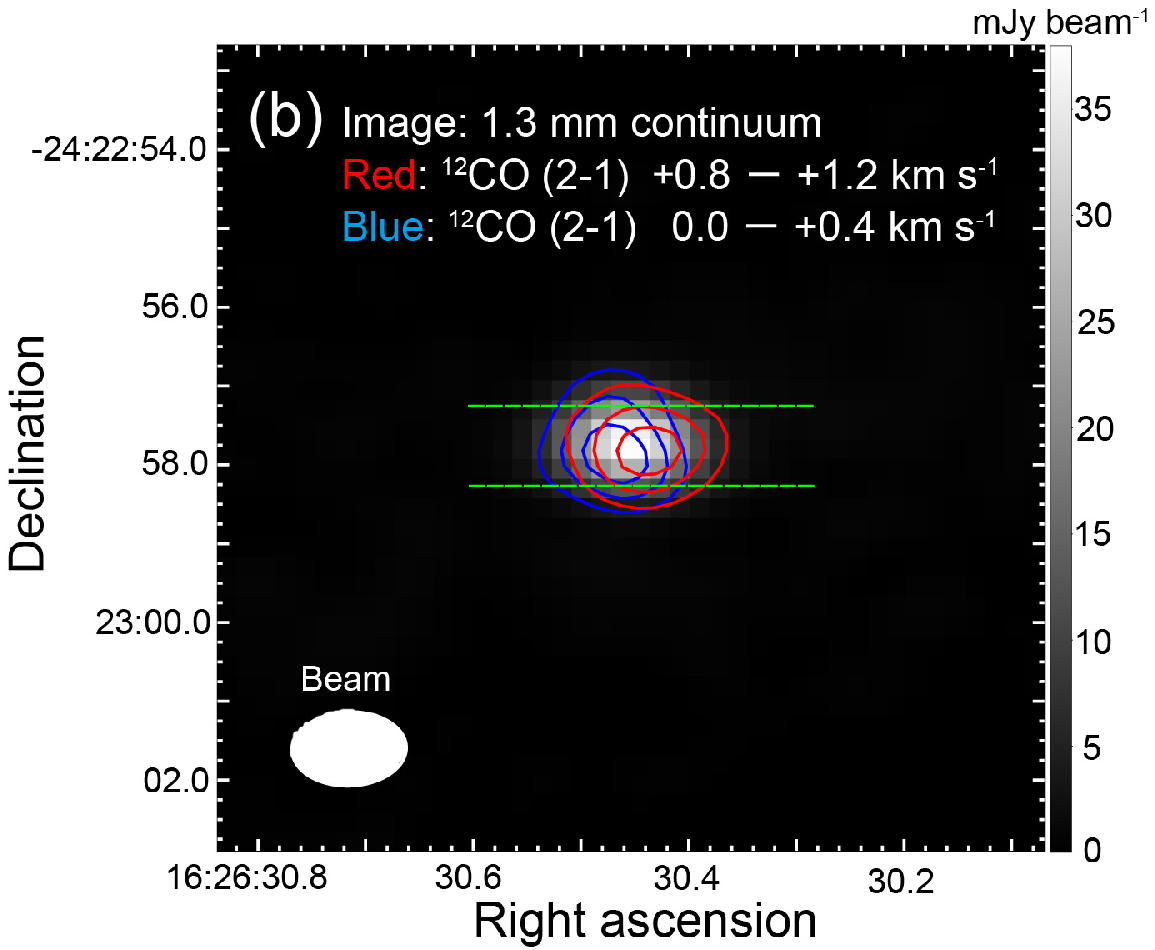}
\epsscale{0.45}
\plotone{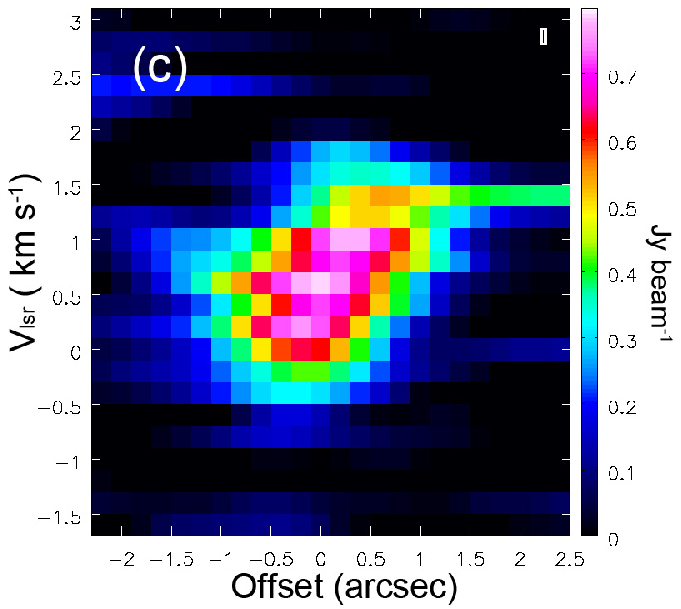}
\caption{(a) $^{12}$CO (2--1), $^{13}$CO (2--1), and C$^{18}$O (2--1) spectra of GY-51 made from a region of $3\farcs6\times2\farcs4$. (b) $^{12}$CO (2--1) contours overlaid on the 1.3 mm continuum map of GY-51. The red (0.8 -- 1.2~km~s$^{-1}$) and blue (0 -- 0.4~km~s$^{-1}$) components are integrated separately on the $^{12}$CO (2--1) map, which have 0.4~km~s$^{-1}$ width symmetrically relative to the peak velocity of 0.6~km~s$^{-1}$ (see spectra in panel a). The contour levels are common in the blue and red components, and are 0.2, 0.3, and 0.4~Jy~beam$^{-1}$~km~s$^{-1}$. The beam size is shown as an ellipse in the bottom left. (c) $^{12}$CO (2--1) PV diagram of GY-51, which is calculated from the pixels between the two green dashed lines in panel b.}
\label{GY-51}
\end{figure}

In the PDR shell, the 1.3~mm continuum map (Fig.~\ref{channel_map}) shows a compact and strong peak shell with total flux density of 40.5~mJy within a $3\farcs6\times2\farcs4$ region (=2.5$\times$beam size), which is spatially coincident with the YSO, GY-51 (\citealt{Bontemps01}).
The spectral index of $\alpha=0.05$ in 2--24~$\micron$ (\citealt{Evans03}) indicates that GY-51 is a flat-spectrum YSO.
Figure \ref{GY-51}a shows $^{12}$CO ($J=$2-1), $^{13}$CO ($J=$2-1), and C$^{18}$O ($J=$2-1) spectra toward GY-51 made from a $3\farcs6\times2\farcs4$ region.
In Fig.~\ref{GY-51}, CO data obtained from only the 12-m Array are presented to examine relatively compact ($<10\arcsec$) structures around GY-51.
In GY-51, the $^{12}$CO emission is clearly detected (3.2$\pm$0.1 Jy km s$^{-1}$), while the $^{13}$CO and C$^{18}$O emission is not detected (3$\sigma$ upper limit: $<0.15$ Jy km s$^{-1}$).
The 3$\sigma$ lower limit of the $I(^{12}{\rm CO}(2-1))/I(^{13}{\rm CO}(2-1))$ intensity ratio is $>21$.
An $I(^{12}{\rm CO}(2-1))/I(^{13}{\rm CO}(2-1))$ intensity ratio measured toward another flat-spectrum YSO, HL Tau, is $\sim$4 (\citealt{Wu18}), which is comparable to that in Class II YSOs (\citealt{Williams14}).
Hence GY-51 has an unusually high $I(^{12}{\rm CO}(2-1))/I(^{13}{\rm CO}(2-1))$ intensity ratio.
Since GY-51 is a flat-spectrum YSO, its remnant envelope might be retained around GY-51.
As shown in Fig.~\ref{GY-51}b, however, no extended emission components are detected in the 1.3~mm continuum.
In addition, a $^{12}$CO ($J=$2-1) position-velocity map (Fig.~\ref{GY-51}c) shows only a rotation-like velocity gradient and no clear sign of an infalling envelope.
$^{12}$CO contours of the blue and red sides are also overlaid on Fig.~\ref{GY-51}b.
Hence, the remnant envelope around GY-51 may be negligible, and the high $I(^{12}{\rm CO}(2-1))/I(^{13}{\rm CO}(2-1))$ intensity ratio may originate from emission from the disk.

The high $I(^{12}{\rm CO}(2-1))/I(^{13}{\rm CO}(2-1))$ intensity ratio is consistent with the CO layered structures detected at the PDR shell, suggesting that the CO isotopologues are affected by the selective dissociation not only in the PDR shell but also in GY-51.
Based on the 1.3~mm dust continuum flux of 40.5~mJy, the dust mass of the disk around GY-51 is estimated as $M_\mathrm{dust} = 1.6\times 10^{-5}M_{\odot}$ where $T_{\mathrm{dust}}=60$~K and $\kappa_{\rm 1.3mm}=2.3$ cm$^2$ g$^{-1}$ are assumed.
If we simply assume that a gas-to-dust mass ratio of 100 (i.e., $M_\mathrm{gas} = 1.6\times 10^{-3}M_{\odot}$), a CO-to-H$_2$ abundance ratio of $6\times 10^{-5}$, a $T_{\mathrm{gas}}=60$ K, and a disk radius of 70 au, the optical depth of $^{12}$CO is roughly estimated to be $\sim 5.2\times 10^2$ (e.g., \citealt{Turner91}).
Therefore, if the $X$($^{12}$CO)/$X$($^{13}$CO) abundance ratio is the typical interstellar value of 62 (\citealt{Langer93}), the optical depth of $^{13}$CO is $\sim$8 and the line should be detectable.

Predicted values of $I(^{12}{\rm CO} (2-1))/I(^{13}{\rm CO} (2-1))$ intensity ratios for disks were calculated by \citet{Williams14} (see also \citealt{Miotello14, Miotello16}) using a wide variety of parameter ranges and taking into account the freeze-out of CO gas on dust grains near the midplane of the outer disk, photo-dissociation at the disk surface, and selective photo-dissociation.
According to the model calculations, the observed gas mass of $\sim 1.6\times 10^{-3}M_{\odot}$ is too large to achieve the high intensity ratio of $>$21.
If the gas mass is so large, $^{13}$CO would also be self-shielded, and then only small portion of the CO gas would be affected by the selective photo-dissociation.
In this case, both the $^{12}$CO and $^{13}$CO lines would be optically thick, and the intensity ratio should be small.
Thus, the high intensity ratio suggests that the disk gas mass is small (i.e., the gas-to-dust mass ratio is smaller than 100) and that large portion of the CO gas is affected by the selective photo-dissociation due to the strong UV radiation ($G_0=(3-4)\times 10^3$; \citealt{Shimajiri17}).
In such a situation, a small gas-to-dust mass ratio could be caused by photo-evaporation of the gas.
Therefore, a combination of the photo-evaporation and the selective photo-dissociation may explain the high $I(^{12}{\rm CO}(2-1))/I(^{13}{\rm CO}(2-1))$ intensity ratio toward GY-51.

\section{Conclusion}

We have analyzed $^{12}$CO(2--1), $^{13}$CO(2--1), C$^{18}$O (2--1), and 1.3~mm continuum maps of $\rho$ Oph A with ALMA.
As a result, layered structures of the CO isotopologues due to the selective dissociation are clearly detected.
The angular separation between the layers is $\sim$10$\arcsec$, which is not spatially resolved using any current single-dish telescopes.
Around the PDR shell, the $X$($^{13}$CO)/$X$(C$^{18}$O) abundance ratio changes from 20--40 to 5.5 in a small angular scale of $\sim$4$\arcsec$ ($2.6\times10^{-3}$~pc), which is reasonably consistent to the plane-parallel PDR model calculation considering the density of $n_\mathrm{H} \sim10^5$~cm$^{-3}$ in the corresponding area.
We have also found an intensity ratio of $I$($^{12}$CO(2--1))/$I$($^{13}$CO(2--1))$>$21 in a flat-spectrum YSO, GY-51, in the $\rho$ Oph A PDR, which is significantly higher than those in flat-spectrum and Class II YSOs ($\sim$4).
These results indicate that UV radiation can significantly affects the CO isotopologues via selective dissociation in the overall $\rho$ Oph A PDR.
In addition, the $\rho$ Oph A PDR is a plane-parallel PDR which is suitable for various PDR studies.

\acknowledgments

We express many thanks to the anonymous referee for useful comments.
This paper makes use of the following ALMA data: ADS/JAO.ALMA\#2013.1.00839.S. ALMA is a partnership of ESO (representing its member states), NSF (USA) and NINS (Japan), together with NRC (Canada) and NSC and ASIAA (Taiwan) and KASI (Republic of Korea), in cooperation with the Republic of Chile. The Joint ALMA Observatory is operated by ESO, AUI/NRAO and NAOJ.
MY and ST acknowledges grants from JSPS KAKENHI Grant Numbers JP17K14261, JP16H07086, and JP18K03703.
YS received support from the ANR (project NIKA2SKY, grant agreement ANR-15-CE31-0017).
This work was supported by NAOJ ALMA Scientific Research Grant Number 2017-04A.

\bibliographystyle{aasjournal}

\begin{thebibliography}{}
\expandafter\ifx\csname natexlab\endcsname\relax\def\natexlab#1{#1}\fi
\providecommand{\url}[1]{\href{#1}{#1}}

\bibitem[{{Anders} \& {Grevesse}(1989)}]{Anders89}
{Anders}, E., \& {Grevesse}, N. 1989, \gca, 53, 197

\bibitem[{{Bontemps} {et~al.}(2001){Bontemps}, {Andr{\'e}}, {Kaas}, {Nordh},
  {Olofsson}, {Huldtgren}, {Abergel}, {Blommaert}, {Boulanger}, {Burgdorf},
  {Cesarsky}, {Cesarsky}, {Copet}, {Davies}, {Falgarone}, {Lagache},
  {Montmerle}, {P{\'e}rault}, {Persi}, {Prusti}, {Puget}, \&
  {Sibille}}]{Bontemps01}
{Bontemps}, S., {Andr{\'e}}, P., {Kaas}, A.~A., {et~al.} 2001, \aap, 372, 173

\bibitem[{{Draine}(1978)}]{Draine78}
{Draine}, B.~T. 1978, \apjs, 36, 595

\bibitem[{{Evans} {et~al.}(2003){Evans}, {Allen}, {Blake}, {Boogert}, {Bourke},
  {Harvey}, {Kessler}, {Koerner}, {Lee}, {Mundy}, {Myers}, {Padgett},
  {Pontoppidan}, {Sargent}, {Stapelfeldt}, {van Dishoeck}, {Young}, \&
  {Young}}]{Evans03}
{Evans}, II, N.~J., {Allen}, L.~E., {Blake}, G.~A., {et~al.} 2003, \pasp, 115,
  965

\bibitem[{{Genzel} {et~al.}(1988){Genzel}, {Harris}, {Jaffe}, \&
  {Stutzki}}]{Genzel88}
{Genzel}, R., {Harris}, A.~I., {Jaffe}, D.~T., \& {Stutzki}, J. 1988, \apj,
  332, 1049

\bibitem[{{Glassgold} {et~al.}(1985){Glassgold}, {Huggins}, \&
  {Langer}}]{Glassgold85}
{Glassgold}, A.~E., {Huggins}, P.~J., \& {Langer}, W.~D. 1985, \apj, 290, 615

\bibitem[{{Goicoechea} {et~al.}(2017){Goicoechea}, {Cuadrado}, {Pety}, {Bron},
  {Black}, {Cernicharo}, {Chapillon}, {Fuente}, \& {Gerin}}]{Goicoechea17}
{Goicoechea}, J.~R., {Cuadrado}, S., {Pety}, J., {et~al.} 2017, \aap, 601, L9

\bibitem[{{Hollenbach} {et~al.}(1991){Hollenbach}, {Takahashi}, \&
  {Tielens}}]{Hollenbach91}
{Hollenbach}, D.~J., {Takahashi}, T., \& {Tielens}, A.~G.~G.~M. 1991, \apj,
  377, 192

\bibitem[{{Kamazaki} {et~al.}(2001){Kamazaki}, {Saito}, {Hirano}, \&
  {Kawabe}}]{Kamazaki01}
{Kamazaki}, T., {Saito}, M., {Hirano}, N., \& {Kawabe}, R. 2001, \apj, 548, 278

\bibitem[{{Kamazaki} {et~al.}(2003){Kamazaki}, {Saito}, {Hirano}, {Umemoto}, \&
  {Kawabe}}]{Kamazaki03}
{Kamazaki}, T., {Saito}, M., {Hirano}, N., {Umemoto}, T., \& {Kawabe}, R. 2003,
  \apj, 584, 357

\bibitem[{{Kawabe} {et~al.}(2018){Kawabe}, {Hara}, {Nakamura}, {Saigo},
  {Kamazaki}, {Shimajiri}, {Tomida}, {Takakuwa}, {Tsuboi}, {Machida}, {Di
  Francesco}, {Friesen}, {Hirano}, {Oasa}, {Tamura}, {Tamura}, {Tsukagoshi}, \&
  {Wilner}}]{Kawabe18}
{Kawabe}, R., {Hara}, C., {Nakamura}, F., {et~al.} 2018, \apj, 866, 141

\bibitem[{{Keene} {et~al.}(1985){Keene}, {Blake}, {Phillips}, {Huggins}, \&
  {Beichman}}]{Keene85}
{Keene}, J., {Blake}, G.~A., {Phillips}, T.~G., {Huggins}, P.~J., \&
  {Beichman}, C.~A. 1985, \apj, 299, 967

\bibitem[{{Langer} \& {Penzias}(1993)}]{Langer93}
{Langer}, W.~D., \& {Penzias}, A.~A. 1993, \apj, 408, 539

\bibitem[{{Larsson} \& {Liseau}(2017)}]{Larsson17}
{Larsson}, B., \& {Liseau}, R. 2017, \aap, 608, A133

\bibitem[{{Lin} {et~al.}(2016){Lin}, {Shimajiri}, {Hara}, {Lai}, {Nakamura},
  {Sugitani}, {Kawabe}, {Kitamura}, {Yoshida}, {Tatei}, {Akashi}, {Higuchi}, \&
  {Tsukagoshi}}]{Lin16}
{Lin}, S.-J., {Shimajiri}, Y., {Hara}, C., {et~al.} 2016, \apj, 826, 193

\bibitem[{{Liseau} {et~al.}(2015){Liseau}, {Larsson}, {Lunttila}, {Olberg},
  {Rydbeck}, {Bergman}, {Justtanont}, {Olofsson}, \& {de Vries}}]{Liseau15}
{Liseau}, R., {Larsson}, B., {Lunttila}, T., {et~al.} 2015, \aap, 578, A131

\bibitem[{{Loren} {et~al.}(1983){Loren}, {Sandqvist}, \& {Wootten}}]{Loren83}
{Loren}, R.~B., {Sandqvist}, A., \& {Wootten}, A. 1983, \apj, 270, 620

\bibitem[{{Mardones} {et~al.}(1997){Mardones}, {Myers}, {Tafalla}, {Wilner},
  {Bachiller}, \& {Garay}}]{Mardones97}
{Mardones}, D., {Myers}, P.~C., {Tafalla}, M., {et~al.} 1997, \apj, 489, 719

\bibitem[{{Meixner} \& {Tielens}(1993)}]{Meixner93}
{Meixner}, M., \& {Tielens}, A.~G.~G.~M. 1993, \apj, 405, 216

\bibitem[{{Miotello} {et~al.}(2014){Miotello}, {Bruderer}, \& {van
  Dishoeck}}]{Miotello14}
{Miotello}, A., {Bruderer}, S., \& {van Dishoeck}, E.~F. 2014, \aap, 572, A96

\bibitem[{{Miotello} {et~al.}(2016){Miotello}, {van Dishoeck}, {Kama}, \&
  {Bruderer}}]{Miotello16}
{Miotello}, A., {van Dishoeck}, E.~F., {Kama}, M., \& {Bruderer}, S. 2016,
  \aap, 594, A85

\bibitem[{{Motte} {et~al.}(1998){Motte}, {Andre}, \& {Neri}}]{Motte98}
{Motte}, F., {Andre}, P., \& {Neri}, R. 1998, \aap, 336, 150

\bibitem[{{Nishimura} {et~al.}(2015){Nishimura}, {Tokuda}, {Kimura}, {Muraoka},
  {Maezawa}, {Ogawa}, {Dobashi}, {Shimoikura}, {Mizuno}, {Fukui}, \&
  {Onishi}}]{Nishimura15}
{Nishimura}, A., {Tokuda}, K., {Kimura}, K., {et~al.} 2015, \apjs, 216, 18

\bibitem[{{Ortiz-Le{\'o}n} {et~al.}(2017){Ortiz-Le{\'o}n}, {Loinard},
  {Kounkel}, {Dzib}, {Mioduszewski}, {Rodr{\'{\i}}guez}, {Torres},
  {Gonz{\'a}lez-L{\'o}pezlira}, {Pech}, {Rivera}, {Hartmann}, {Boden}, {Evans},
  {Brice{\~n}o}, {Tobin}, {Galli}, \& {Gudehus}}]{Ortiz-Leon17}
{Ortiz-Le{\'o}n}, G.~N., {Loinard}, L., {Kounkel}, M.~A., {et~al.} 2017, \apj,
  834, 141

\bibitem[{{Paron} {et~al.}(2018){Paron}, {Areal}, \& {Ortega}}]{Paron18}
{Paron}, S., {Areal}, M.~B., \& {Ortega}, M.~E. 2018, \aap, 617, A14

\bibitem[{{Shimajiri} {et~al.}(2013){Shimajiri}, {Sakai}, {Tsukagoshi},
  {Kitamura}, {Momose}, {Saito}, {Oshima}, {Kohno}, \& {Kawabe}}]{Shimajiri13}
{Shimajiri}, Y., {Sakai}, T., {Tsukagoshi}, T., {et~al.} 2013, \apjl, 774, L20

\bibitem[{{Shimajiri} {et~al.}(2014){Shimajiri}, {Kitamura}, {Saito}, {Momose},
  {Nakamura}, {Dobashi}, {Shimoikura}, {Nishitani}, {Yamabi}, {Hara},
  {Katakura}, {Tsukagoshi}, {Tanaka}, \& {Kawabe}}]{Shimajiri14}
{Shimajiri}, Y., {Kitamura}, Y., {Saito}, M., {et~al.} 2014, \aap, 564, A68

\bibitem[{{Shimajiri} {et~al.}(2017){Shimajiri}, {Andr{\'e}}, {Braine},
  {K{\"o}nyves}, {Schneider}, {Bontemps}, {Ladjelate}, {Roy}, {Gao}, \&
  {Chen}}]{Shimajiri17}
{Shimajiri}, Y., {Andr{\'e}}, P., {Braine}, J., {et~al.} 2017, \aap, 604, A74

\bibitem[{{Spaans}(1996)}]{Spaans96}
{Spaans}, M. 1996, \aap, 307, 271

\bibitem[{{Stutzki} {et~al.}(1988){Stutzki}, {Stacey}, {Genzel}, {Harris},
  {Jaffe}, \& {Lugten}}]{Stutzki88}
{Stutzki}, J., {Stacey}, G.~J., {Genzel}, R., {et~al.} 1988, \apj, 332, 379

\bibitem[{{Tauber} {et~al.}(1995){Tauber}, {Lis}, {Keene}, {Schilke}, \&
  {Buettgenbach}}]{Tauber95}
{Tauber}, J.~A., {Lis}, D.~C., {Keene}, J., {Schilke}, P., \& {Buettgenbach},
  T.~H. 1995, \aap, 297, 567

\bibitem[{{Tielens} \& {Hollenbach}(1985)}]{Tielens85}
{Tielens}, A.~G.~G.~M., \& {Hollenbach}, D. 1985, \apj, 291, 722

\bibitem[{{Turner}(1991)}]{Turner91}
{Turner}, B.~E. 1991, \apjs, 76, 617

\bibitem[{{van Dishoeck} \& {Black}(1988)}]{vanDishoeck88}
{van Dishoeck}, E.~F., \& {Black}, J.~H. 1988, \apj, 334, 771

\bibitem[{{Visser} {et~al.}(2009){Visser}, {van Dishoeck}, \&
  {Black}}]{Visser09}
{Visser}, R., {van Dishoeck}, E.~F., \& {Black}, J.~H. 2009, \aap, 503, 323

\bibitem[{{White} {et~al.}(2015){White}, {Drabek-Maunder}, {Rosolowsky},
  {Ward-Thompson}, {Davis}, {Gregson}, {Hatchell}, {Etxaluze}, {Stickler},
  {Buckle}, {Johnstone}, {Friesen}, {Sadavoy}, {Natt}, {Currie}, {Richer},
  {Pattle}, {Spaans}, {di Francesco}, \& {Hogerheijde}}]{White15}
{White}, G.~J., {Drabek-Maunder}, E., {Rosolowsky}, E., {et~al.} 2015, \mnras,
  447, 1996

\bibitem[{{Williams} \& {Best}(2014)}]{Williams14}
{Williams}, J.~P., \& {Best}, W.~M.~J. 2014, \apj, 788, 59

\bibitem[{{Wu} {et~al.}(2018){Wu}, {Hirano}, {Takakuwa}, {Yen}, \&
  {Aso}}]{Wu18}
{Wu}, C.-J., {Hirano}, N., {Takakuwa}, S., {Yen}, H.-W., \& {Aso}, Y. 2018,
  \apj, 869, 59

\bibitem[{{Yurimoto} \& {Kuramoto}(2004)}]{Yurimoto04}
{Yurimoto}, H., \& {Kuramoto}, K. 2004, Science, 305, 1763

\end{thebibliography}

\end{document}